\documentclass[superscriptaddress,letterpaper,twocolumn,prl,showpacs]{revtex4-1}
\usepackage{amsmath, amsthm, amssymb}
\usepackage{graphicx}
\usepackage{dcolumn} 
\usepackage{bm} 

\usepackage{hyperref}
\hypersetup{backref,  
colorlinks=true,
allcolors=blue}

\makeatletter 
\gdef\@ptsize{2}
\let\@currsize\normalsize 
\makeatother 
\usepackage{setspace} 

\usepackage[usenames,dvipsnames]{color}

\begin{document}
\title{Revealing the Hidden Heavy Fermi Liquid in CaRuO$_{3}$}

\author{Yang Liu}
\affiliation{Center for Correlated Matter and Department of Physics, Zhejiang University, China}
\affiliation{Laboratory of Atomic and Solid State Physics, Department of Physics, Cornell University, Ithaca, New York 14853, USA}
\affiliation{Department of Materials Science and Engineering, Cornell University, Ithaca, New York 14853, USA}

\author{Hari P. Nair}
\affiliation{Department of Materials Science and Engineering, Cornell University, Ithaca, New York 14853, USA}

\author{Jacob P. Ruf}
\affiliation{Laboratory of Atomic and Solid State Physics, Department of Physics, Cornell University, Ithaca, New York 14853, USA}

\author{Darrell G. Schlom}  
\affiliation{Department of Materials Science and Engineering, Cornell University, Ithaca, New York 14853, USA}
\affiliation{Kavli Institute at Cornell for Nanoscale Science, Ithaca, New York 14853, USA}

\author{Kyle M. Shen}
\email[Yang Liu and Hari P. Nair contribute equallly to this work. Author to whom correspondence should be addressed: ]{kmshen@cornell.edu}
\affiliation{Laboratory of Atomic and Solid State Physics, Department of Physics, Cornell University, Ithaca, New York 14853, USA}
\affiliation{Kavli Institute at Cornell for Nanoscale Science, Ithaca, New York 14853, USA}



\begin{abstract}
The perovskite ruthenate has attracted considerable interest due to reports of possible non-Fermi-liquid behavior and its proximity to a magnetic quantum critical point, yet its ground state and electronic structure remain enigmatic. Here we report the first measurements of the Fermi surface and quasiparticle dispersion in CaRuO$_{3}$ through a combination of oxide molecular beam epitaxy and \emph{in situ} angle-resolved photoemission spectroscopy. Our results reveal a complex and anisotropic Fermi surface consisting of small electron pockets and straight segments, consistent with the bulk orthorhombic crystal structure with large octahedral rotations. We observe a strongly band-dependent mass renormalization, with prominent heavy quasiparticle bands which lie close to the Fermi energy and exhibit strong temperature dependence. These results are consistent with a heavy Fermi liquid with a complex Fermiology and small hybridization gaps near the Fermi energy. Our results provide a unified framework for explaining previous experimental results on CaRuO$_{3}$, such as its unusual optical conductivity, and demonstrate the importance of octahedral rotations in determining the quasiparticle band structure, and electron correlations in complex transition metal oxides. 
\end{abstract}


\maketitle


The ruthenates host a remarkably diverse class of exotic quantum phases, ranging from spin-triplet superconductivity, ferromagnetism, metamagnetism, spin-density waves, and quantum criticality, all with the same basic building block of corner-sharing RuO$_{6}$ octahedra with a central Ru$^{4+}$ ion \cite{MackenzieRMP2003}\cite{GrigeraScience2001}\cite{KosterRMP2012}. Amongst the ruthenates, CaRuO$_{3}$ remains a particularly enigmatic compound. Measurements of the optical conductivity ($\sigma_{1} \propto \omega^{-1/2}$) and resistivity ($\rho(T) \propto T^{3/2}$) have suggested that paramagnetic CaRuO$_{3}$ exhibits a non-Fermi liquid (NFL) ground state \cite{LeePRB2002}\cite{KleinPRB1999}\cite{CapognaPRL2002}\cite{CaoSSC2008}, where the electronic excitations cannot be mapped directly to single-electron excitations, giving rise to physical properties not described by conventional Fermi liquid (FL) theory. Indeed, given its close similarity to its isostructural and isoelectronic ferromagnetic counterpart, SrRuO$_3$, it has been argued that CaRuO$_3$ might be on the cusp of a magnetic quantum critical point \cite{KleinPRB1999}\cite{CaoSSC2008}\cite{MazinPRB1997}, given the strong ferromagnetic fluctuations seen in nuclear magnetic resonance and induced ferromagnetism by defects and dopings \cite{YoshimuraPRL1999}\cite{HePRB2001}\cite{MaignanPRB2006}\cite{DurairajPRB2006}. On the other hand, the strong interplay between Hund's coupling $J$ and electronic onsite repulsion $U$ in CaRuO$_{3}$ could give rise to a fragile FL with a low coherence temperature, as recently proposed by dynamical mean-field theory (DMFT) \cite{GeorgesARCMP2013}\cite{MravljePRL2011}\cite{DangPRL2015}\cite{DangPRB2015}\cite{DengPRL2016} and supported by transport measurements below 2 K \cite{SchneiderRPL2014}. In particular, it has been proposed theoretically that  the large RuO$_{6}$ octahedral rotations in CaRuO$_{3}$ may give rise to a multitude of low lying interband transitions that could mimic NFL effects in the optical conductivity \cite{DangPRL2015}\cite{GeigerPRB2015}. Nevertheless, precise knowledge of the momentum-dependent electronic structure, particularly the Fermi surface (FS), is crucial for understanding the true nature of the ground state and electromagnetic properties of CaRuO$_{3}$.

In this Rapid Communication, we report the first momentum-resolved measurements of quasiparticle (QP) dispersions and FS in CaRuO$_{3}$, by combining high quality thin film growth by reactive-oxide molecular-beam epitaxy (MBE) and \emph{in situ} angle-resolved photoemission spectroscopy (ARPES) measurements. Our data reveal sharp, well-defined QP excitations that form a complex band structure arising from the large GdFeO$_{3}$-type distortions in CaRuO$_{3}$, confirming its FL ground state. We observe a manifold of heavy, flat QP bands close to the Fermi energy ($E_{F}$) caused by large octahedral rotations. Our measurement of the low-energy electronic structure provides a unified framework for explaining both the unconventional optical and terahertz conductivity \cite{LeePRB2002}\cite{DangPRL2015}\cite{SchneiderRPL2014}\cite{GeigerPRB2015} as arising from low-lying interband transitions, as well as the large electronic specific heat ($\sim$80 mJ / mol K$^{2}$) \cite{CaoSSC2008}\cite{KikugawaJPSJ2009}, and crossover behavior in resistivity and the Hall coefficient with temperature, originating from the unexpectedly heavy QP bands.

\begin{figure}[tb]
	\includegraphics[width=0.5\textwidth]{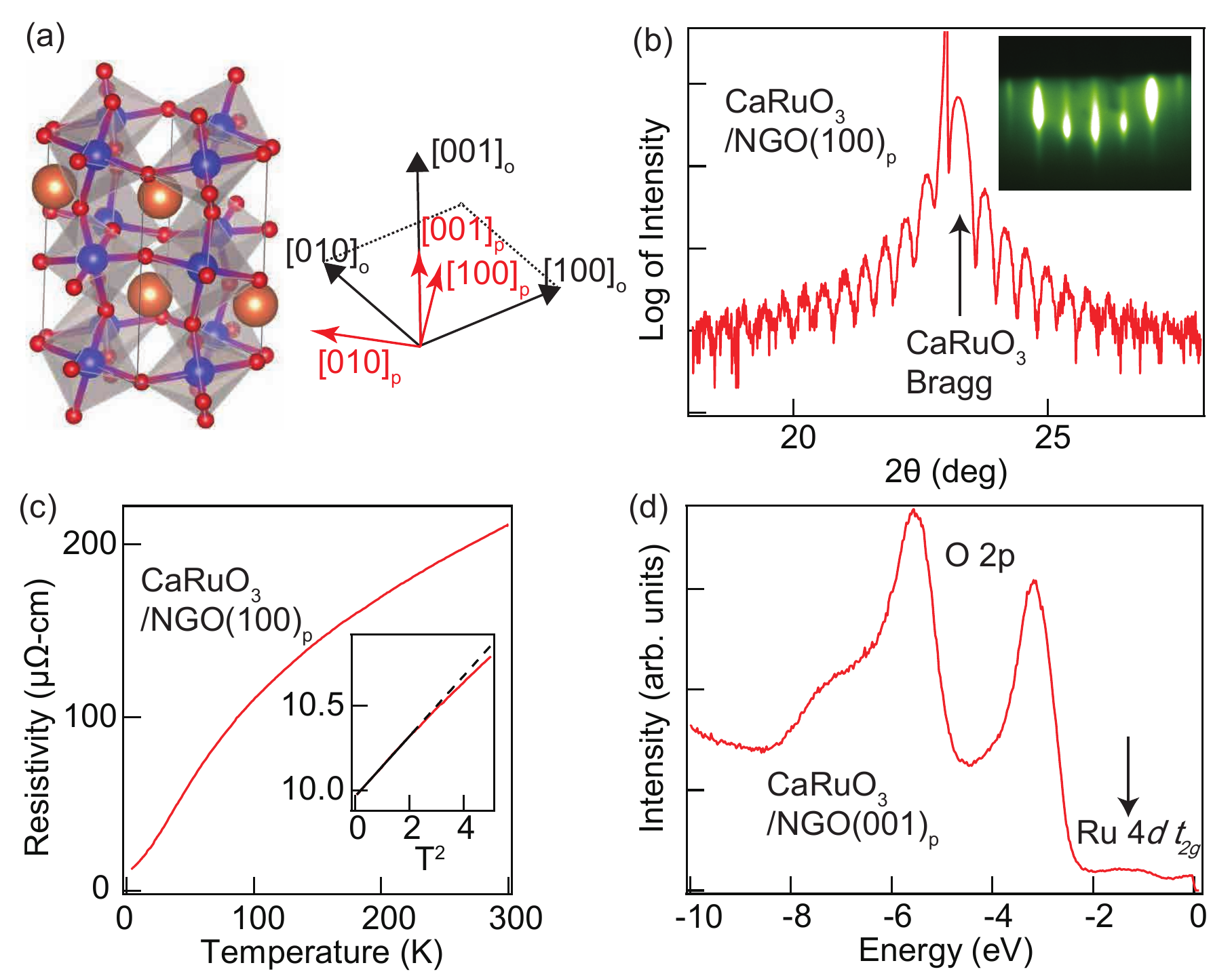}
	\caption{\label{fig:SpectralFunctionEvolution} (a) Crystal structure (the orthorhombic unit cell is indicated by black box) and correspondence between orthorhombic ($_{o}$) and pseudocubic ($_{p}$) lattice. The brown, blue, red spheres represent Ca, Ru, O atoms, respectively. (b) \emph{ex situ} X-ray diffraction scan along the out-of-plane direction. Inset shows the RHEED pattern along the $<$001$>_{p}$ azimuth angle during the growth. (c) Resistivity as a function of temperature for the same ARPES sample shown in Fig. 2(d-f). Inset is a $\rho$ vs $T^{2}$ plot at very low temperature, highlighting a very low FL coherence temperature of $\sim1.5$ K, deduced by deviation from a simple $T^{2}$ fitting at low temperature (black dashed line). (d) Valence band photoemission along the surface normal direction for CaRuO$_{3}$ (001)$_{p}$ film. }
\end{figure}


Epitaxial thin films of CaRuO$_{3}$, typically $\sim20$ nm, were grown by MBE in a dual-chamber Veeco GEN10 system in an oxidant ($\sim90\%$ O$_{2}$ +  $10\%$ O$_{3}$) background pressure of 8 $\times$ 10$^{-7}$ torr and a substrate temperature of 800$^{\circ}$ C, as measured using the k-space BandiT detector operating in blackbody mode. The film growth was monitored in real time using reflection high-energy electron diffraction (RHEED). Growth of $(001)$ and $(110)$-oriented CaRuO$_{3}$ films were achieved by selecting similarly oriented NdGaO$_{3}$ (NGO) substrates. Immediately after growth, thin films were transferred under ultrahigh vacuum to a high-resolution ARPES system consisting of a VG Scienta R4000 analyzer and a VUV5000 helium plasma discharge lamp and monochromator \cite{MonkmanNM2012}. Measurements were performed at 17 K (unless noted otherwise) with an energy resolution $\Delta E = 10$ meV with He I$\alpha$ ($h\nu = 21.2$ eV) photons and a base pressure of 7 $\times 10^{-11}$ torr. Spectra were also taken with He II ($h\nu = 40.81$ eV) photons to confirm the bulk nature of the observed bands shown in the paper. After ARPES measurements, samples were characterized in detail by \emph{in situ} low-energy electron diffraction (LEED), \emph{ex situ} x-ray diffraction, electrical transport, and Hall measurements.


Figure 1(a) shows the crystal structure of bulk CaRuO$_{3}$ with its $a^{-}a^{-}c^{+}$ type octahedral rotations \cite{GlazerAC2012} which cause highly distorted RuO$_{6}$ octahedra and an orthorhombic lattice with lattice constants close to ($\sqrt{2} a_{0}, \sqrt{2} a_{0}, 2a_{0}$), where $a_{0}$ is the pseudocubic lattice constant. Also shown in Fig. 1(a) is the correspondence between the orthorhombic ($_{o}$) and pseudocubic ($_{p}$) coordinate systems, which is commonly used to highlight the effect of octahedral rotations. The well-defined thickness fringes in x-ray diffraction (Fig. 1(b)) demonstrate the epitaxial, single-phase, and atomically flat nature of the thin films. In addition, RHEED (Fig. 1(b)) and low-energy electron diffraction (LEED, Fig. 2(a,d)) confirm that films with both the (001)$_{o}$  (same as (001)$_{p}$) and (110)$_{o}$ (same as (100)$_{p}$) out-of-plane orientations can be stabilized, which have subsequently been confirmed by \emph{ex situ} transmission electron microscopy and synchrotron x-ray diffraction measurements. The typical residual resistivity ratios (RRRs = $\rho(300K)/\rho(4K)$) of samples measured by ARPES are on the order of 20, and RRRs on other samples as high as 75 have been measured \cite{NairAPLMaterials2018}, indicating the high quality of the films. The resistivity exhibits a FL-like $\rho \propto T^{2}$ behavior below $\sim1.5$ K (hence a fragile FL), and cross over to a $\rho \propto T^{1.5}$ behavior above 2 K, consistent with previous measurements \cite{KleinPRB1999}\cite{CapognaPRL2002}\cite{CaoSSC2008}\cite{SchneiderRPL2014}\cite{KikugawaJPSJ2009}. Measurements of the valence band (Fig. 1(d)) show the O $2p$ and Ru $4d$ states, consistent with earlier reports by Yang \emph{et al.} which focused on the origin of the broad hump around 1 eV binding energy (marked by an arrow) as being due to enhanced correlations \cite{YangPRB2016n2}.


\begin{figure}[tb]
	\includegraphics[width=0.5\textwidth]{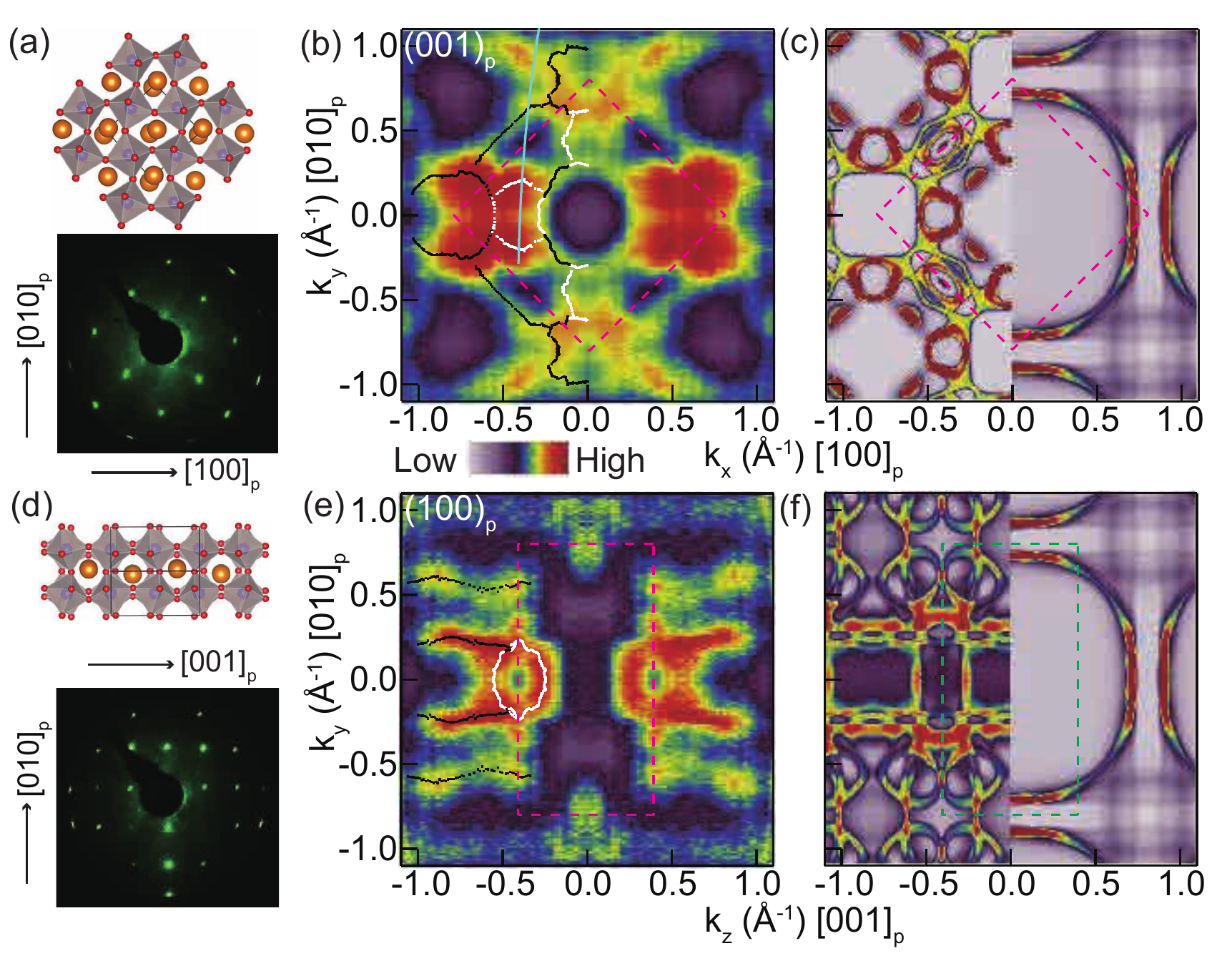}
	\caption{\label{fig:SpectralFunctionEvolution} (a) and (d) top view of the crystal structure of (001)$_{p}$ (a) and (100)$_{p}$ (d) terminated surface, showing a ($\sqrt{2}$x$\sqrt{2}$)$R45^{\circ}$ and 2x2 LEED pattern in pseudocubic coordinate, respectively (taken at 66 eV). (b) and (e) are the experimentally measured FSs of CaRuO$_{3}$ (001)$_{p}$ (b) and (100)$_{p}$ (e) films, which are presented in image plots with color scale shown in the middle (red indicates high intensity, light blue indicate low intensity). The purple dashed lines in (b,c,e,f) mark the projected bulk Brillouin zone (BZ) boundary. The extracted Fermi contours are displayed on the left half as black and white dots, for light and heavy QPs, respectively. The slanted cyan line in (b) shows the momentum cut position in Fig. 3. (c) and (f) are the simulated ARPES spectra from DFT calculations for (001)$_{p}$ (c) and (100)$_{p}$ (f) films, respecively. Left halves in (c,f) are simulations using a realistic orthorhombic structure, and right halves are simulations assuming a cubic structure without octahedral rotation. The simulations of the FS maps use appropriate $k_{z}$ to match the experimental data and assume a $k_{z}$ broadening (0.1$\pi/a_{0}$) and lifetime broadening (quadratic in energy expected for a FL), as described in details in \cite{Supplemental}. }
\end{figure}

In Fig. 2, we show maps in momentum space of the ARPES intensity at $E_{F} \pm 10$ meV, for films aligned both along the (001)$_{p}$ (Fig. 2(b))and (100)$_{p}$ (Fig. 2(e)) directions. Both orientations show a complex FS comprised of small pockets arising from the large GdFeO$_{3}$-type distortions, which cause band folding and hence small hole or electron pockets. The experimental Fermi wavevectors ($k_{F}$s) from maxima in either the momentum distribution curves (MDCs) or energy distribution curves (EDCs) are summarized in the left halves in Fig. 2(b,e) as black dots. For the (001)$_{p}$ films, the experimental FS exhibits small pockets centered at (0,0), ($\pi/2,0$) and ($\pi,0$), and possibly ($\pi/2,\pi/2$) (all defined within the pseudocubic coordinate). In comparison, we also plot DFT simulations  of the corresponding $k$-space maps calculated using Wien2K in the generalized gradient approximation (details in Supplemental Materials \cite{Supplemental}) in Figs. 2(c,f): left for the bulk structure (structure parameters adapted from \cite{ZayakPRB2006}), right for the ideal cubic structure with 180$^{\circ}$ Ru-O-Ru bonds. By comparing the experimental data with various $k_{z}$ slices through the DFT calculation, we estimate $k_{z} \sim 0.5 \pi/a_{0}$ for the (001)$_{p}$ surface and $k_{z} \sim 0.7 \pi/a_{0}$ for the (100)$_{p}$ surface under $h\nu = 21.2$ eV photons (Figs. S1 and S2 in \cite{Supplemental}). Those values of $k_{z}$ suggests an inner potential of $14.4 \pm 2$ eV and $11 \pm 2$ eV for the (001)$_{p}$ and (100)$_{p}$ surface respectively, similar to SrRuO$_{3}$ \cite{Takizawathesis2007}. The orthorhombic DFT calculations in the left half of Fig. 2(c) qualitatively reproduces the multiplicity of small FS pockets, in contrast to the case of SrRuO$_{3}$, where the band folding is much weaker \cite{MackenziePRB1998}\cite{AlexanderPRB2005}\cite{ShaiPRL2013}\cite{YangPRB2016}. Despite sharing the same $Pbnm$ structure, the rotation angles of the oxygen octahedra are approximately doubled in CaRuO$_{3}$ compared to SrRuO$_{3}$ (the averaged rotation angle along each of the three pseudocubic axis is $\sim 5^{\circ}$ in SrRuO$_{3}$ versus $\sim 11^{\circ}$ for CaRuO$_{3}$ \cite{ChengPNAS2013} \cite{KiyamaPRB1996}). This leads to a significant difference in the momentum distribution of spectral weight, from reflecting nearly cubic symmetry in SrRuO$_{3}$, to a much more complex structure in CaRuO$_{3}$, and could be important for the observed differences in electromagnetic properties (Table I in \cite{Supplemental}). 

The importance of octahedral rotations is even more evident when comparing the (001)$_{p}$ data to that from the (100)$_{p}$ surface which should be identical in the idealized cubic structure without rotations (right halves in Fig. 2(c,f)). The ARPES Fermi surface maps show dramatic differences (Fig. 2(e)) between the two orientations. For instance, in the (100)$_{p}$ films, there is only a single enclosed pocket near ($\pi/2,0$) together with long straight segments of high intensity running parallel to the [001]$_{p}$ direction which qualitatively match the corresponding orthorhombic DFT simulations for this surface. The average radius of this electron pocket is measured to be 0.15 \AA$^{-1}$, which is in agreement with the only frequency seen from SdH oscillations (0.12 \AA$^{-1}$) in similarly oriented (100)$_{p}$ films \cite{SchneiderRPL2014}. SdH oscillation results from an (001)$_{p}$ film are not available at the moment, but additional frequencies have been observed as the magnetic field is moved from [100]$_{p}$ towards the [001]$_{p}$ direction \cite{SchneiderRPL2014}, which is consistent with the ARPES FS map of (001)$_{p}$ films, showing additional small pockets due to strong band folding.   


\begin{figure}[tb]
	\includegraphics[width=0.5\textwidth]{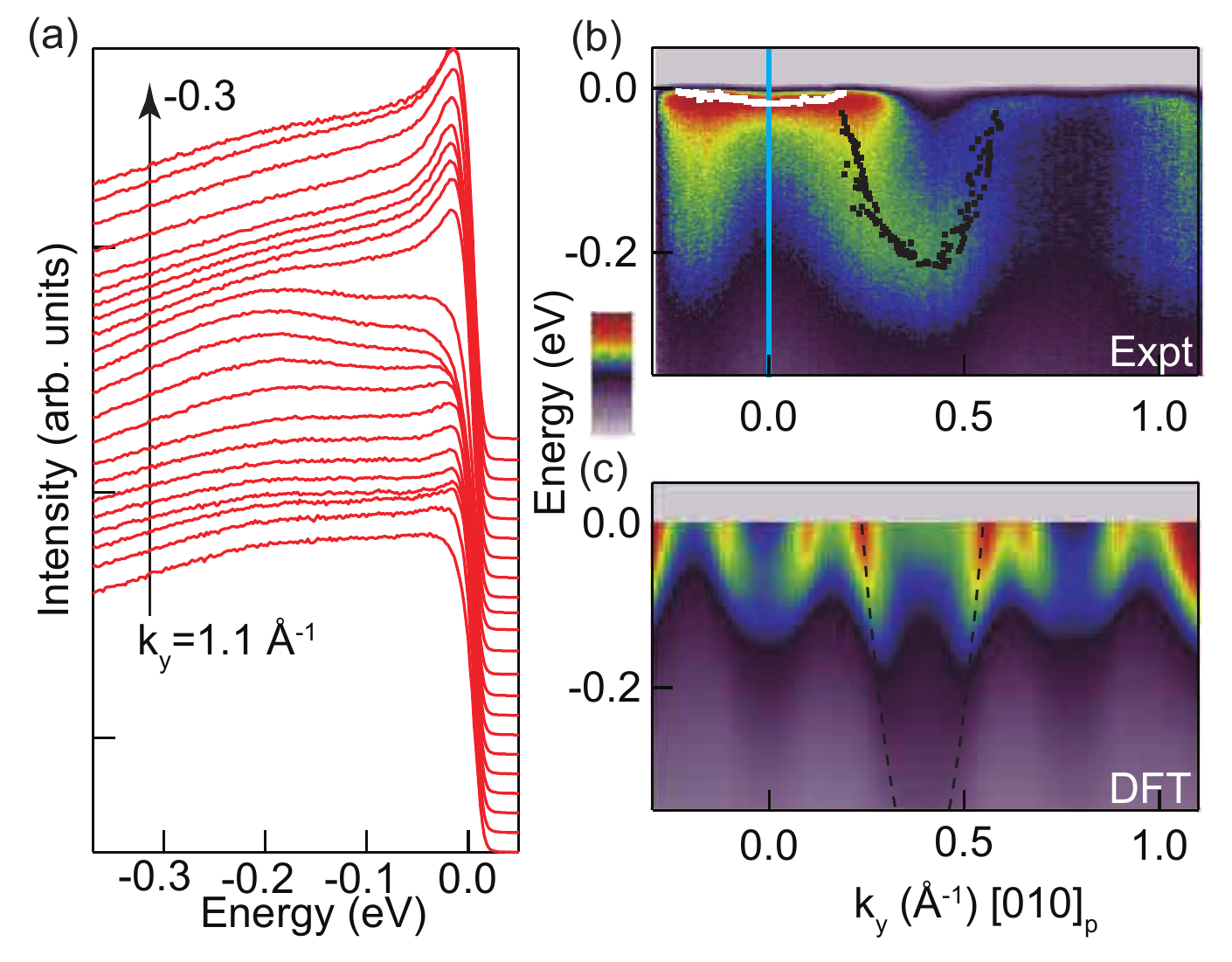}
	\caption{\label{fig:mdc}Waterfall (a) and image plot (b) of energy-momentum cuts for (001)$_{p}$ oriented films. The cut is along the slanted cyan line shown in Fig. 2(b). The experimentally extracted QP dispersion is shown as white (black) dots, for heavy (light) QP bands, respectively. The heavy QP dispersion near $E_{F}$ is extracted from Fig. S5 in \cite{Supplemental}. (c) shows simulated ARPES spectra based on DFT calculations, along the same momentum cut as (b). Black dashed curves indicate DFT calculated light QP dispersion. Appropriate energy and k$_{z}$ broadening have been taken into account for the simulation \cite{Supplemental}. }  
\end{figure}

In Fig. 3(a), we show the EDCs along the momentum cut shown by the cyan line in Fig. 2(b). A weakly dispersive, sharp QP peak is clearly observed close to $E_{F}$ and whose intensity is highly dependent on the RRR, underscoring the importance of sample quality. The image plot is shown in Fig. 3(b), which also reveals a neighboring, broader band with significant dispersion (effective mass 0.8 $m_{e}$). To accurately extract the heavy QP dispersion, we divide the measured ARPES spectra by the Fermi-Dirac function convolved by a Gaussian resolution function \cite{Supplemental}. The results reveal that the heavy QP band is electron-like with a fitted effective mass $m^{\ast}$ = 13.5 $\pm$ 1.5 $m_{e}$ (black dots in Fig. 3(b)). The experimental data on (100)$_{p}$ films further confirm that these heavy QPs possess large masses along all three momentum directions (Fig. S4 in \cite{Supplemental}), which allows us to calculate the electronic specific heat associated with these heavy QPs to be 60 $\pm$ 6.7 mJ / mol K$^{2}$ \cite{Supplemental}, accounting for a large portion of the experimental specific heat ($\sim$80 mJ / mol K$^{2}$). The remainder could be due to contributions from other lighter bands. DFT calculations for the bulk-like structure predicts $m^{\ast} \sim 1 m_{e}$ for all relevant bands (see Fig. 3(c)) \cite{Supplemental}. This strongly band-dependent renormalization, and the subsequent coexistence of heavy and light QPs near $E_{F}$ is remarkable and the origin of such large variations in the renormalization of bands from presumably the same orbitals remains unclear. As a result, only the light QP bands measured in experiment agree well with the DFT simulations of the Fermi surface and band dispersion (Figs. 2c, 2f, 3c), while the heavy QP bands show strong discrepancies with the DFT calculations due to their much stronger renormalization (the heavy bands were not used in comparing the Fermi surface topologies or $k_{z}$ determination for this reason). This strongly band-dependent renormalization and major discrepancies for the heavy bands indicate that DFT alone cannot explain the QP dispersion in CaRuO$_{3}$, and the inclusion of onsite Hubbard repulsion and Hund's coupling is likely essential. A full description of the electronic structure requires advanced theoretical tools, such as DFT + DMFT \cite{GeorgesARCMP2013}\cite{DangPRB2015}\cite{DengPRL2016}.

We emphasize that the observed QP bands are most likely derived from bulk, rather than surface states, given the clear differences in data taken with different photon energies (Fig. S6 in \cite{Supplemental}). Moreover, the  heat capacity estimated from ARPES measurements matches closely to the bulk thermodynamic measurements, suggesting that the heavy QP bands which dominate this calculation probably correspond to bulk electronic states. In addition, the light QP bands are in qualitative  agreement with bulk DFT calculations. While it is difficult to conclusively rule out subtle surface relaxations, the combination of the photon energy dependence and the agreement with bulk measurements and calculations, when taken as a whole, suggests that any possible surface relaxation is minimal enough not to qualitatively affect our observations, as supported by the LEED measurements which are also consistent with the bulk symmetry (Fig. 2).

A central question for CaRuO$_{3}$ remains the robustness of the FL ground state, given the apparently low FL coherence temperature (1.5 K) deduced from resistivity measurements (Fig. 1(c)). We perform temperature dependent ARPES measurements and in fact observe well-defined QP peaks up to 100 K (Fig. 4(a)), thus providing direct spectroscopic evidence of robust FL-like QPs even to high temperatures. A detailed lineshape analysis shows the disappearance of the heavy QPs at $\sim150$ K in Fig. 4(b), corresponding to the crossover in resistivity (Fig. 1(c)) and rapid change of Hall coefficient \cite{Supplemental}\cite{GausepohlPRB1996}. Such a temperature dependence has also been reported in other ruthenates \cite{ShimoyamadaPRL2009}\cite{KondoPRL2016} and has been proposed to be a direct signature of strong correlations in a Hund's metal \cite{MravljePRL2011}, where the interplay between $U$ and $J$ results in a large intermediate temperature range where the coherent spectral weight shows strong temperature dependence \cite{DangPRB2015}\cite{XuPRL2013}\cite{DengPRL2013}. The observation of heavy QPs with a strong temperature dependence, together with the large electronic specific heat and crossover behavior in resistivity and Hall coefficient, indicate a heavy Fermi liquid ground state with surprising similarities to heavy fermion systems \cite{SteglichPRL1979}. Whether there is any inherent connection between two systems would be an interesting topic of investigation for future studies.


\begin{figure}[tb]
	\includegraphics[width=0.5\textwidth]{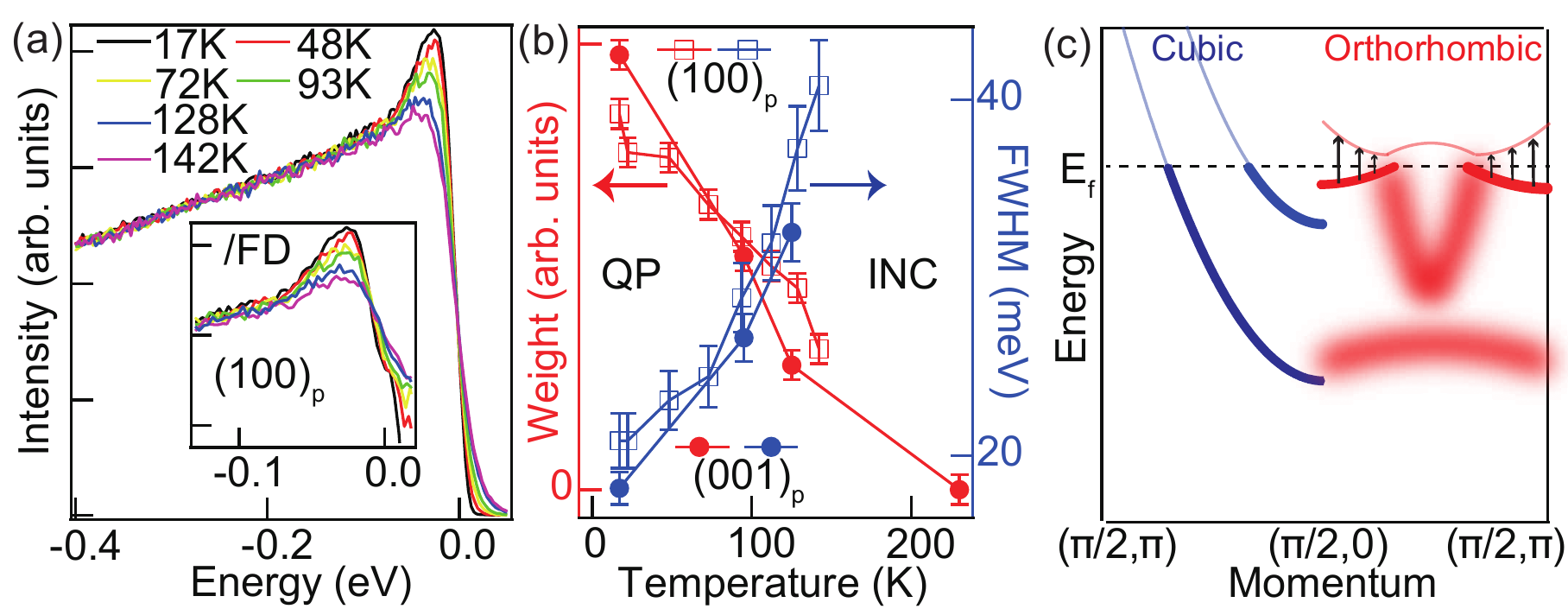}
	\caption{\label{fig:mdc}(a) Temperature dependence of heavy QP band for (100)$_{p}$ film at the ($\pi/2,0$) pocket (white vertical cut in Fig. S4(b) in \cite{Supplemental}). The inset is the EDCs divided by Fermi-Dirac function convolted by a Gaussian function, revealing the full spectral function above $E_{F}$. (b) Deduced QP spectral weight (red symbols, left axis) and full width at half maximum (FWHM, blue symbols, right axis), which is proportional to the QP scattering rate. Data from both a (001)$_{p}$ film (closed circle) and a (100)$_{p}$ film (open rectangular) are shown. The charge carriers exhibit gradual crossover from well-defined QP at low temperature to incoherent (INC) carrier at high temperature. The raw data for the (001)$_{p}$ film is included in Fig. S5(b) in \cite{Supplemental}. (c) Cartoon showing the origin and implication of heavy QP in CaRuO$_{3}$. The horizontal (vertical) axis is the momentum (energy). Left half illustrates the QP spectra for the cubic phase without rotation, and the right half shows the case afer considering rotation.}  
\end{figure}

In Fig. 4(c), we show a schematic summarizing our key observations for CaRuO$_{3}$, where large octahedral rotations cause strong band folding and hybridization, resulting in a complex Fermi surface topology with many small electron or hole pockets and hybridization gaps near $E_{F}$. Our results allows for a unified understanding of not only its electronic structure, but also the myriad of experimental observations previously reported in the literature. The complex band structure, comprised of many heavy bands which lie within 30 meV (7.25 THz or 242 cm$^{-1}$) of $E_{F}$, is the origin of the multitude of low-energy interband transitions, which mimics the signature of NFL optical conductivity previously reported. The large rotations also significantly reduce the bandwidth, leading to large mass renormalization for some bands near $E_{F}$, while other light QPs coexist. This strongly band-dependent mass renormalization is remarkable and its origin is yet to be understood theoretically. The heavy QPs exhibit strong temperature dependence, a signature of strong correlation in Hund's metals. Our results not only provide a first complete experimental understanding of the complex electronic structure of CaRuO$_{3}$, but generally highlights the importance of octahedral rotations in correlated ruthenate perovskites and how it can impact the fermiology and physical properties, in a much more pronounced manner than other prototypical metallic perovskites, e.g., SrVO$_{3}$/CaVO$_{3}$ \cite{YoshidaPRB2010}.

We would like to thank A. Georges, Q. Han, and A.J. Millis for very helpful discussions and insights. This work was supported by the National Science Foundation through DMR-0847385, DMR-1308089, and through the Materials Research Science and Engineering Centers program (DMR-1719875, the Cornell Center for Materials Research), and the Research Corporation for Science Advancement (2002S). This work was performed in part at the Cornell NanoScale Facility, a member of the National Nanotechnology Infrastructure Network, which is supported by the National Science Foundation (Grant No. ECCS-0335765). Y. L. acknowledges support from National Natural Science Foundation of China (Grant No. 11674280) and National key R$\&$D program of the MOST of China (Grant No.2016YFA0300203).


\end{document}